\documentclass[journal]{IEEEtran}
\usepackage[utf8]{inputenc}
\inputencoding{utf8}
\usepackage{graphicx}

\usepackage[english]{babel}
\usepackage{listings}
\usepackage{fancyvrb}
\usepackage{framed}
\usepackage[listings,skins]{tcolorbox}
\usepackage[skipbelow=\topskip,skipabove=\topskip]{mdframed}
\mdfsetup{roundcorner=0}

\usepackage{graphicx}
\usepackage{enumitem}
\usepackage{authblk}
\usepackage[T1]{fontenc}

\ifCLASSINFOpdf
\else
\fi

\title{%
A Publish/Subscribe QoS-aware Framework for Massive IoT Traffic Orchestration}
\author{Pedro F. D. Moraes\thanks{pedro@omoraes.com.br}}
\author[2]{Rafael F. Reale\thanks{reale@ifba.edu.br}}
\author{Joberto S. B. Martins\thanks{joberto.martins@unifacs.br }}

\affil{Salvador University - UNIFACS}
\affil[2]{Instituto Federal da Bahia - IFBA}

\begin{document}
\maketitle

\selectlanguage{english}
\begin{abstract}
Internet of Things (IoT) application deployment requires the allocation of resources such as virtual machines, storage, and network elements that must be deployed over distinct infrastructures such as cloud computing, Cloud of Things (CoT), datacenters and backbone networks. For massive IoT data acquisition, a gateway-based data aggregation approach is commonly used featuring sensor / actuator seamless access and providing cache / buffering and preprocessing functionalities. In this perspective, gateways acting as producers need to allocate network resources to send IoT data to consumers. In this paper, it is proposed a Publish/Subscribe (Pub/Sub) quality of service (QoS) aware framework (PSIoT-Orch) that orchestrates IoT traffic and allocates network resources between aggregates and consumers for massive IoT traffic. PSIoT-Orch framework schedules IoT data flows based on its configured QoS requirements. Additionally,  the framework allocates network resources (LSP - bandwidth) over a controlled backbone network with limited and constrained resources between IoT data users and consumers. Network resources are allocated using a Bandwidth Allocation Model (BAM) to achieve efficient network resource allocation for scheduled IoT data streams. The framework adopts an ICN (Information-Centric Network) Pub/Sub architecture approach to handle IoT data transfers requests among framework components. The proposed framework aims at gathering the inherent advantages of an ICN-centric approach using a PubSub message scheme while allocating resources efficiently keeping QoS awareness and handling restricted network resources (bandwidth) for massive IoT traffic.  
%
%
%
%
%
%
\end{abstract}


%
\IEEEpeerreviewmaketitle

\section{Introduction}
%
%
%
%
\IEEEPARstart{T}{he} Internet of Things (IoT) is considered an important trend in many areas like smart cities, smart grid, e-health, industry and future Internet  \cite{These-Elie} \cite{borgia2014}. As such, a large effort is being undertaken to find suitable technologies, standards, middlewares and architectures to support IoT application deployment.  IoT deployment typically requires the orchestration of heterogeneous resources that are allocated over distinct infrastructures such as cloud computing, cloud of things (CoT), datacenters and backbone networks \cite{These-Elie}.  

IoT devices like sensors and actuators are in large quantity for most IoT applications and, in addition, they differ considerably in terms of processing, storage and functional capabilities. In addition, IoT  setups generate a huge amount of heterogeneous traffic leading to an increasing quality of service (QoS), resource allocation and network configuration complexity.

Paradigms such as Cloud Computing \cite{cloud}, and more recently Fog Computing \cite{fog}, arise to alleviate the weight that massive IoT data processing and traffic has on networks and devices. In particular Fog Computing relies on mostly operating on the edges of the network, to minimize the load on the whole network by serving already processed and aggregated data. However, traffic still might need to be sent across the network to interested clients, and as the amount of IoT devices increases and spreads geographically, processed IoT traffic served by IoT/ Fog-like aggregators along the edges might still heavily load the network if no specific traffic management is done.

In an attempt to address this scenario, we present a Pub/ Sub QoS-aware framework (PSIoT-Orch) for managing massive IoT traffic aggregated into Fog-like IoT gateways along the network edge. This framework allows for IoT quality of service traffic management according to network-wide specifications, application domain and IoT characteristics. Aspects like the backbone network topology, network traffic saturation and IoT domain requirements are considered.

In this article we'll describe the PSIoT-Orch framework  components, its relation to IoT requirements and how they are combined to seamlessly manage IoT traffic. In section \ref{sssec:iotfeatures} we explore proposed IoT-oriented architectures and proposals and how they pertains to our framework. Section \ref{sssec:frameworkintro} is an overview of the framework components as well as key features and purpose. In Section \ref{sssec:proof-of-concept} we describe how we've built our simple proof-of-concept implementation and network evaluation scenario. Finally, section \ref{sssec:conclusion} concludes with an overview of what has been achieved.

\section{Key IoT Aspects and Related Work} \label{sssec:iotfeatures}

IoT has potential for the creation of new intelligent applications in nearly every field, with it's devices that enable local or mobile sensing and actuation services.

The different fields of application can be organized in different ways into various domains like industrial, smart city and health among others \cite{borgia2014}. For every standardized or practical field, IoT devices share common features and requirements in traffic and usage. Our framework builds on dealing with distinct IoT requirements, namely heterogeneity, scalability and QoS.

While IoT devices, traffic characteristics and requirements are quite well defined, architectures for IoT generally have a hard time maintaining interoperability with each other. In 2009 the ETSI Technical Committee for Machine-to-Machine communications (ETSI TC M2M) was established to develop a reference IP-based architecture relying on existing technologies \cite{etsi}. This architecture has three domains: a) the Application Domain, where client and M2M applications reside; b) the Network Domain, consisting of any network between applications and device gateways; c) the Device \& Gateway domain, where all the devices and/or gateways are located.

Our framework fits well in this sort of architecture, building a bridge in the Network Domain between device gateways and applications.

One approach for massive scale IoT data dissemination is presented in \cite{daneels2017}. In this proposal, remote and rural areas are the focus and an ad-hoc interconnection infrastructure is adopted for IoT traffic transport using a mix of low-power  wireless personal area network (LoWPAN) and wireless sensor network (WSN). To the best of our knowledge, the communications resources between IoT aggregators and consumer applications mostly use cloud computing, fog-like services, Internet connectivity, ad-hoc solutions or a mix of them \cite{These-Elie} \cite{fog}. Our proposal, distinctly from commonly used approaches, uses a controlled network with limited resources in which bandwidth utilization and optimization is the focus. 

\section{PSIoT-Orch Framework QoS Awareness, Network Resource Allocation and Pub/Sub Orchestration} \label{sssec:frameworkintro}

The overall goal of the proposed PSIoT-Orch framework is to manage the massive traffic generated by a huge number of IoT devices, aiming to handle efficiently network resources and IoT QoS requirements over the network between the IoT devices and consumer IoT applications. Consumers might be hosts over the backbone network, cloud computing infrastructures accessed by the network or any other scheme that makes use of the managed network infrastructure for communication (Figure 1).

IoT gateways (IoTGW-Ag) act as traffic aggregators interacting with IoT devices. IoT gateway traffic aggregators use a publish/subscribe style architecture lined along the network edge to transmit their data to application clients (consumers), with these transmissions being mediated by a centralized orchestrator  (PSIoT-Orchestrator) (Figure 1).

\begin{figure}[!htb]
\includegraphics[width=\columnwidth]{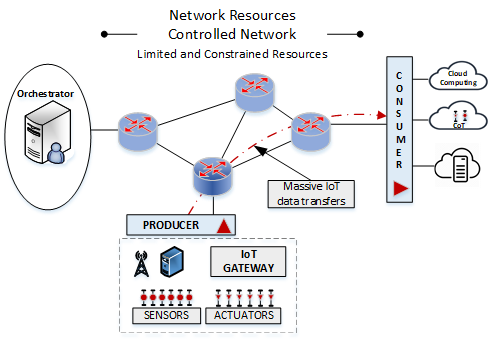}
\caption{PSIoT-Orch framework basic components}
\label{Framework-Components}
\end{figure}

\subsection{QoS-aware IoT Traffic Resource Deployment}

The PSIoT-Orch framework offers a set of QoS levels based on time-sensitivity to deal with IoT traffic requirements regarding time-sensitivity for massive data transmission: 
%
%
%
%

\begin{itemize}
\item Insensitive: best-effort transmission (e.g., weather data, non-critical smart home data).
\item Sensitive: low data transmission delay (e.g., commercial data, security sensors).
\item Priority: high transmission rate and low delay (e.g., health care, critical industrial sensors).
\end{itemize}

The PSIoT-Orch QoS levels are intended to allow the creation of IoT traffic classes (TCs) over the backbone. These time-sensitivity classes group IoT traffic having similar application requirements and allows a QoS-aware arbitration of IoT massive data flows over the backbone network. The traffic scheduling is done for these defined classes and will require the allocation of network resources. 

As indicated, subsequently and independently, PSIoT-Orch operation will deploy network communication resources (LSPs, bandwidth) based on IoT traffic classes (TCs). As such, applications' IoT data flows are simultaneously prioritized by the QoS scheduler and communication resources are efficiently allocated  over the network using a bandwidth allocation model based strategy (BAM-based).

\subsection{Network Resource Allocation based on Bandwidth Allocation Model (BAM)}
Once IoT traffic is scheduled according with its QoS requirements, a further aspect mediated by the PSIoT-Orch framework is the effective network resource allocation (LSP; bandwidth) between producer and consumer(s).

The architectural approach adopted by the framework is to delegate the allocation of bandwidth resources to a bandwidth allocation model (BAM).

Using BAM to allocate resources is an implementation option that is aligned with the fact that the backbone network resources availability is assumed to be restricted. This a real scenario of typical backbone setups interconnecting IoT devices through either private, metropolitan or long-distance networks (like smart cities or large distributed IoT setups). BAM, in turn, have a proven capability to distribute and manage efficiently scarce network resources \cite{GBAM}. Consequently, relying on BAMs as a broker-like entity to allocate bandwidth is a relevant option that must be considered for the PSIoT-Orch framework. This is specially valid when the massive IoT traffic generated by producers will somehow reduce and affect the overall network resources availability and bandwidth dispute will occur.

\subsection{PSIoT-Orch Pub/Sub Message Scheme as an Information-Centric Network (ICN)}
PSIoT-Orch framework uses a Publish/ Subscribe (Pub/Sub) Information-Centric Network (ICN) message scheme in which an information-based access model is used \cite{ICN1}. As such, consumers try to access named content (IoT data) without any direct mapping to the transport mechanism used over the network that interconnect them with data producers.

This message scheme provides the inherent ICN advantages to the framework including dynamic discovery and dissemination, efficient distribution of content, the potential to factorize functionalities and in-aggregator cache that may save energy and increase local content availability \cite{icn-baccelli} \cite{sinc2016} \cite{An}.

\section{PSIoT-Orch Framework} \label{sssec:psiotframework}

PSIoT-Orch framework has 4 main components that interact to provide network resource allocation with quality of service awareness (Figure \ref{Framework-Components}):

\begin{itemize}
\item IoT gateways data aggregators (IoTGW-Ag)  acting as "Producers" and IoT applications acting as "Consumers";
\item The "PSIoT-Orchestrator" acting as the mediator; and
\item The backbone network interconnecting efficiently the previous components.
\end{itemize}

In addition, the PSIoT-Orch framework functional operation is composed by 2 distinct and cooperating functional entities:

\begin{itemize}
\item The Pub/Sub message scheme allowing the asynchronous request of IoT data over the entire infrastructure; and 
\item The quality of service (QoS) and network resource allocation police and implementation scheme provided by  the "PSIoT-Orchestrator".
\end{itemize}

In sequence, these framework architectural components, functional entities and overall framework overview are described. 

\subsection{PSIoT-Orch Framework Overview}

The PSIoT-Orch frameworks main role is to manage and monitor the traffic output from each IoT aggregator according to the overall aggregators throughput and the underlying network resource availability and usage (saturation). 

In PSIoT-Orch, each IoT aggregator works in a similar paradigm to Fog Computing. Each aggregator node is lined up along the network edge collecting IoT traffic from local devices and offering them to applications via a topic-based publish/subscribe interface. While PSIoT-Orch framework only manage traffic aggregation, it can also be further extended to perform Fog-like capabilities such as IoT raw data processing and manipulation.

\subsection{PSIoT-Orch Gateway Aggregator}

\begin{figure}[!htb]
\includegraphics[width=\columnwidth]{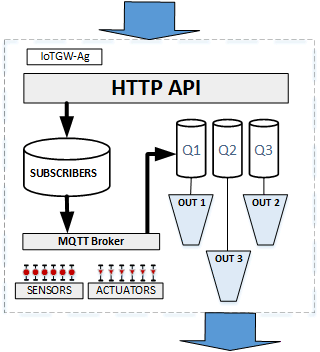}
\caption{IoTGW-Ag aggregator internal structure}
\label{IoTGWAg}
\end{figure}

Each PSiOT-Orch gateway aggregator (IoTGW-Ag) can be sectioned into two main roles: a) that of gathering IoT traffic from local devices and of b) sending consumers their subscribed data, observing the transmission effort each QoS level should follow, as defined by the orchestrator (Figure \ref{IoTGWAg}).

The IoTGW-Ag section that deals with the actual gathering of IoT data from devices can be based on either MQTT (Message Queue Telemetry Transport) technology \cite{mqtt}, traffic generators for simulated devices or other generalized data gateways that interface with the IoTGW-Ag. All these options must preserve the topic-based nature of the framework’s subscription data.

The network-facing section of the IoTGW-Ag deals with consumer topic requests, via a HTTP publish/subscribe API, sending the orchestrator the metadata related to the IoTGW-Ag’s own subscriptions, transmission rates and buffer state, as well as keeping track of the orchestrator-defined transmission rates for each QOS level.

\subsection{The Pub/Sub QoS Configuration Message Scheme}

\begin{figure}[!htb]
\includegraphics[width=\columnwidth]{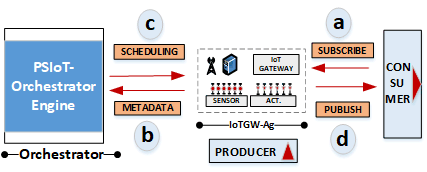}
\caption{PSIoT-Orch Pub/Sub messages}
\label{PubSubMessages1}
\end{figure}

Figure \ref{PubSubMessages1} shows the initial flow of configuration messages between subscribers, providers and orchestrator allowing the IoT data transfers:

\begin{enumerate}[label=(\alph*)]
\item A consumer initiates a topic subscription, with a requested QoS level, to a specific IoTGW-Ag
\item The IoTGW-Ag sends to the orchestrator relevant metadata such as number of subscribers and their QoS levels and buffer allocation.
\item The orchestrator notifies the IoTGW-Ag with an amount of bandwidth that can be consumed by each level of QoS (TCs).
\item The IoTGW-Ag publishes the data to the client according to bandwidth and data availability in the buffer.
\end{enumerate}

As IoTGW-Ags receive applications topic subscriptions, they must notify the orchestrator so as to maintain updated the level of information the orchestrator needs to manage for each aggregators data output. While the framework's orchestration isn't distributed, the failure of the orchestration component would not entail in the failure of the IoTGw-Ag data delivery, since they maintain the set of predefined output rates in the case of an orchestrator failure occurs.

\subsection{The Pub/Sub Network Resource Allocation Message Scheme }

Once IoT data has been published with its attributed QoS level, consumers may request them. At this point, a second level of intervention by the PSIoT-Orchestrator  is necessary to create an effective communication channel (LSP/ bandwidth)  between consumer(s) and producer over their interconnection backbone network.

Figure \ref{PubSubMessages1} details the flow of information exchange between any IoTGW-Ag and the orchestrator. This communication is asynchronous and subscribing applications are not aware of this, nor is their subscription dependent on the communication with the orchestrator. Using a IoTGW-Ag that uses an MQTT broker to aggregate IoT data from devices in the local network, this process will be described in a linear fashion:

\begin{enumerate}
\item Consumers subscribe to any given IoTGW-Ag, requesting their desired topics and QoS levels. Consumers must be aware of the three QoS levels available in the framework.
\item Internally, the topic subscriptions will be recorded in the IoTGW-Ag's list of active subscriptions.
\item Subscriptions are registered in the IoTGW-Ag's internal IoT MQTT broker.
\item Sensor data is published from IoT devices to the IoTGW-Ag's internal IoT MQTT broker.
\item The subscribed data is passed on to the IoTGW-Ag's buffers with different QoS levels, according to each topic subscribers requested QoS level.
\item The IoTGw-Ag buffers store the messages that should be published to subscribed applications.
\item The IoTGW-Ag requests the allocation of communications resource to the PSIoT-Orchestrator and a LSP (Label Switched Path - LSP) is allocated (or not) considering an efficient distribution of the available network resources.
\item Each buffer is emptied (IoT data transmitted to the application) by using the communication channel (LSP, other) allocated by the BAM model and consistent with its QoS level as determined by the Orchestrator.
\end{enumerate}

\begin{figure}[!htb]
\includegraphics[width=\columnwidth]{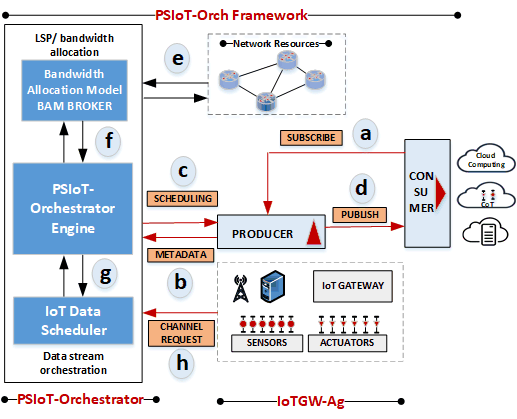}
\caption{PSIoT-Orch framework overall components message exchange}
\label{PubSubMessages22}
\end{figure}

Other operational aspects of the data exchange between IoTGW-Ag and the orchestrator are:

\begin{itemize}
\item Each IoTGW-Ag should periodically (necessarily on each new subscription) send the metadata that the orchestrator needs for traffic scheduling (e.g., buffer sizes, message frequency, subscriber list, topics, QoS levels).
\item The orchestrator then decides the effort with which each level of QoS must publish the data to the applications. This decision will depend on several factors, such as the state of saturation of the network and priority of other IoTGW-Ag according to each IoTGW-Ag's reported metadata. 
%
%
\item Once the decision is made, the orchestrator sends each IoTGW-Ag the transmission effort that each level of QOS must have. The frequency each IoTGW-Ag's receives transmission effort changes is dependent on the algorithm that the orchestrator is relying on to decide the transmission rates.
\end{itemize}

\subsection{PSIoT-Orchestrator}

The PSiOT-Orchestrator has two main components: the QoS IoT traffic scheduler module and the bandwidth allocation model module (BAM module).

The QoS IoT traffic scheduler module continuously computes the amount of bandwidth that will be associated for each IoT QoS class. The orchestrator functions in a reactive manner: each time it receives an IoTGW-Ag's metadata it will recalculate all IoTGW-Ag's QoS level transmission rates. This is the default behavior, but as the amount of updates becomes a processing burden, the transmission rates calculations can be scheduled to appropriate intervals, according to each orchestrator implementation. The traffic scheduler module's main responsibility is to manage the available bandwidth and distribute it to all IoTGW-Ag's, according to their throughput and QoS-level.

For the orchestrator to be able to properly manage the QoS-aware transmission of all aggregators, it's vital that it maintains updated knowledge on several pertinent factors. These factors include timely updates on aggregator metadata, like subscriber count, buffer usage and throughput.

It's also important to note that aggregators might chose different buffer overflow algorithms according to the IoT device data being collected. For real time sensor data it might not make sense to send outdated data that sits in the aggregator output buffer, so a circular buffer strategy might be more useful in maintaining updated data than a simple first in, first out algorithm.

The bandwidth allocation model module acts as a broker for the available resources on the backbone network (channels). In effect, each time the IoTGW-Ag needs to send data to consumers, it sends the metadata to the PSIoT orchestrator, which in turn requests a communication channel (LSP for a MPLS-aware network) to the PSIoT BAM broker module (message exchange in "f") (Figure \ref{PubSubMessages22}). The request is treated by the BAM module that will grant or deny the channel request, the PSIoT orchestrator will then act accordingly. For computing bandwidth availability, the BAM module will interact with network resources as any BAM implementation does. Information about topology, used bandwidth, current LSPs and other relevant data will be maintained by the BAM module.

The behavior of BAM models has been extensively evaluated and detailed on \cite{DSWITCHING} \cite{BAMReconfiguration}.  For the massive IoT traffic scenario it is assumed that bandwidth dispute will occur between QoS classes (high and low priorities) as far as the network is assumed to have limited resources . For this scenario, the GBAM (Generalized Bandwidth Allocation Model) with AllocTC-Sharing (ATCS) behavior is evaluated in \cite{GBAM}  and is the adopted BAM model in the implementation setup.  ATCS behavior allows resource sharing  between high and low priority traffic classes and provides the best possible network utilization \cite{AllocTCSharing}.

\section{PSIoT-Orch proof-of-concept} \label{sssec:proof-of-concept}
As a proof-of-concept we have developed the three core components of our proposed framework: an orchestrator, an aggregator (producer) and a client (consumer).

We have chosen the network emulator Mininet \cite{mininet} for its SDN capabilities and flexibility in creating network topologies. Mininet is a network emulator that uses lightweight virtualization to run many different network components in a single system. With Mininet we built a simple bandwidth-constrained network with consumers, aggregators, an orchestrator and two hosts to generate network traffic to simulate on-IoT network traffic.

The orchestrator is implemented in a multi-purpose messaging platform, with capabilities similar to publish/subscribe systems. Each aggregator (producer) then connects itself to the orchestrator and receives the designated efforts for each QoS level.

A simple fixed scheduler was used to determine each QoS level for all aggregators (producers). The transmission rates for each QoS level are static and proportionally divided between QoS levels, according to importance. These QoS levels are specific to our producer software and deal directly with each aggregators IoT traffic throughput. 

Because we are mostly interested in the evaluating network traffic, each aggregator (producer) was built to simulate IoT sensor/device data. A configurable rate, topics and data can be attributed to each deployed aggregator (producer). The aggregator is built as a HTTP publish/subscribe server, with none of the usual deliverable guarantees, as those aren’t the focus of the proof-of-concept scenario.

For consumers, another HTTP server is used.  It logs the received IoT data with the ability to configure subscriptions to different aggregators (producers). With this basic setup we can configure different producer-consumer topologies with varied amounts of producers and consumers. The proof-of-concept setup may also include non-IoT traffic between nodes to measure its impact on network congestion and near congestion operation.

\subsection{Proof-of-concept network topology} \label{sssec:mininet-topo}

\begin{figure}[!htb]
\includegraphics[width=\columnwidth]{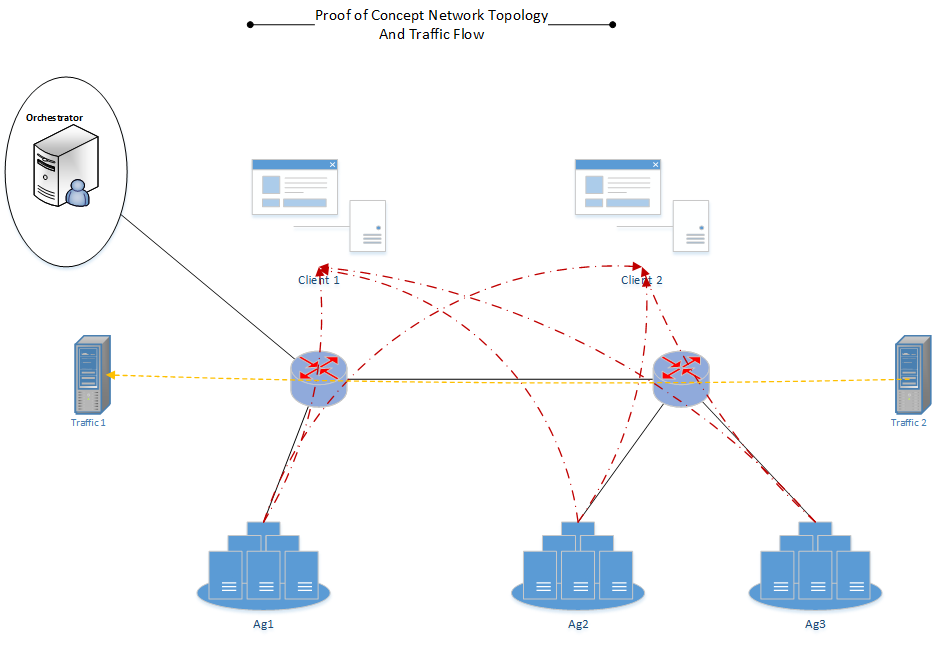}
\caption{Proof-of-concept network topology with two IoT clients and three aggregators}
\label{minitopo}
\end{figure}

In Figure \ref{minitopo} we show the proof-of-concept network topology used, including the main traffic flows expected: from consumers to producers (in red); and between traffic generators (in yellow). Traffic generating hosts are positioned in a way to directly compete with IoT data consumers and producers for network resources (LSP - bandwidth).
To build our network scenario, we created Mininet hosts for each producer, consumer, the orchestrator and non-IoT traffic generators connected with available stock Mininet Switches and Controller, connected by 1MB/s links.
For the orchestrator, we used a simple algorithm to evenly divide the available bandwidth between each aggregator, and for the QoS levels 0, 1 and 2 the individual aggregators bandwidth was allocated in 25\%, 35\% and 45\%, respectively. To accommodate differences in QoS level output, our algorithm also listens for aggregators buffer levels and attempts to allocate more of the available bandwidth accordingly.
Each aggregator then simulates 3 different topics, to generate IoT data, with a reasonable throughput, with Ag1 having a fourth topic that has 100 times more throughput. For our scenario, we follow a timeline, with the following main events:
\begin{enumerate}
\item{Hosts startup: turning on each aggregator, the orchestrator and the non-IoT traffic generators.}
\item{Client 1 initiates subscriptions: Client 1 subscribes to the three IoT topics, one from each aggregator.}
\item{Client 2 initiates subscriptions: Client 2 subscribes to the three IoT topics, one from each aggregator, plus the fourth topic from Ag1.}
\item{End of scenario}
\end{enumerate}

\subsection{Results} \label{sssec:mininet-result}

\begin{figure}[!htb]
\includegraphics[width=\columnwidth]{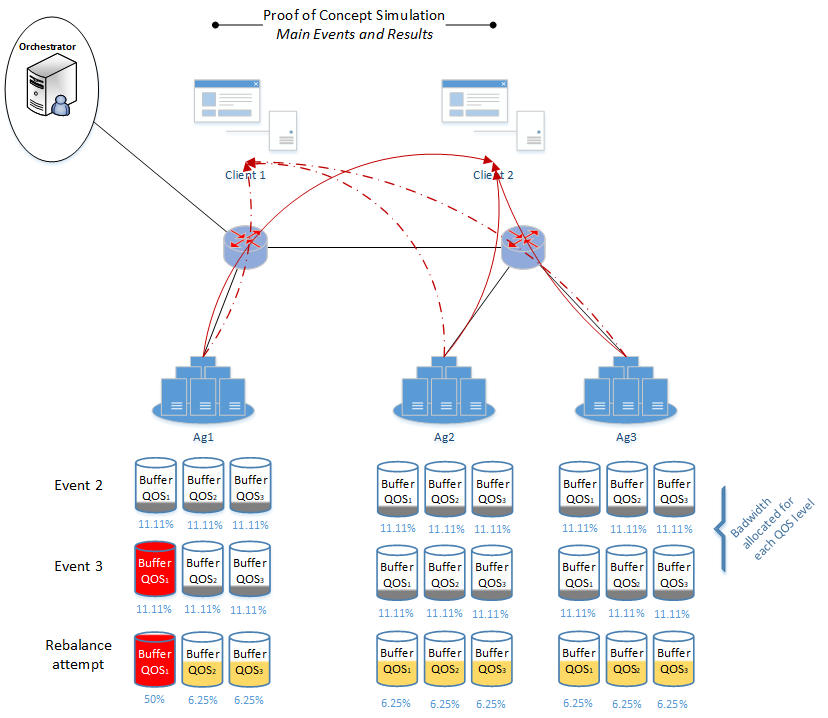}
\caption{Proof-of-concept simulation main events results}
\label{pocbuff}
\end{figure}

After running the scenario, we were able to observe the communication between consumers and producers as well as the management of each aggregator by the orchestrator. After each subscription, the aggregators would report them to the orchestrator, and continue to report their buffer state for each QoS level. Figure \ref{pocbuff} shows the buffer states for the aggregators as well as the allocated bandwidth by the orchestrator for each QOS level, according to each of the main events.s
After the first subscription event, by Client 1, the buffers where constantly at near-empty levels, as the throughput was low enough that the default allocation from the orchestrator was enough. But as soon as Client 2 initiates it’s subscription, along with the fourth topic in Ag1, the buffer in Ag1 begins to fill rapidly.
As soon as the orchestrator detects this, by means of a defined buffer size threshold, it attempts to allocate more bandwidth to Ag1. The orchestrator allocates as much bandwidth that it can without impairing the other subscriptions, but as the data generation rate from Ag1’s fourth topic is much larger than the available bandwidth, we still observe Ag1’s buffer growing as the orchestrator maintains its attempts to allocate enough bandwidth.

\section{Conclusion} \label{sssec:conclusion}

The Internet of Things has arrived and with it the possibility that the massive amount of heterogeneous traffic it can generate will heavily load the backbone networks between producers and consumers of IoT traffic. This article presented the Pub/Sub  PSIoT-Orch framework aiming for massive IoT traffic orchestration relying on IoT QoS needs and efficient management of network resources. 

The emulation-based proof-of-concept demonstrated the basic functionality and that our framework is a viable solution for managing the transmission of massive IoT traffic, as it predictively manages and distributes the available bandwidth between producers. 

The framework proved to be flexible as in both network flow management, but also in maintaining the IoT traffic’s characteristics by transparently offering the same topic-based IoT data from devices to consumers in the network. Furthermore, the ability to have the orchestrator work in tandem with CDN QoS functionalities, allows our framework to be used both integrated to the network, or simply as a 3rd party traffic control between IoT data producers and consumers.

The user-centric approach adopted by the PubSub message scheme between consumers and producers presents an easy interface for data subscription by hiding the complex QoS considerations and data management from both producers and consumers. This allows for consumers to focus on receiving the data and also enables the producers to deal with their Fog-like data aggregation and/or processing.

In terms of future work, it will be developed a real test setup using an network for experimentation testbed (NfExp) distributed in various physical locations  (FIBRE Network) to further validate the flexibility of the framework and simulate BAM behavior over a real producer/consumer IoT backbone.

%

\bibliography{all}
\bibliographystyle{unsrt}

%








\end{document}